\documentclass[aps,prl,twocolumn,pacs]{revtex4}
\usepackage{amsmath,bm,epsfig}
\newcommand{\B}[1]{{\bm{#1}}}

\begin{document}
\title{Dynamic Failure in Amorphous Solids via a Cavitation Instability}
\author{Eran Bouchbinder, Ting-Shek Lo and Itamar Procaccia}
\affiliation{Department of Chemical Physics, Weizmann Institute of Science, Rehovot 76100, Israel}

\date{\today}
\begin{abstract}
The understanding of dynamic failure in amorphous materials via the propagation of free boundaries like cracks and voids  must go beyond elasticity theory, since plasticity intervenes in a crucial and poorly understood manner near the moving free boundary. In this Letter we focus on failure via a cavitation instability in a radially-symmetric stressed material, set up the free boundary dynamics taking both elasticity and visco-plasticity into account, using the recently proposed athermal Shear Transformation Zone theory. We demonstrate the existence (in amorphous systems) of fast cavitation modes accompanied by extensive plastic deformations and discuss the revealed physics.
\end{abstract}
\maketitle

The principal difficulty in describing the mechanical failure of materials using elasticity theory is
that this approach does not provide a law of motion for the free boundary that naturally occurs
when a crack or a void is propagating. Recently, phenomenological phase-field models \cite{phase} were proposed in order to overcome this fundamental
problem. In reality the high stress concentration near the moving
boundary must be associated with some form of plastic deformations that are rather poorly understood in the context of amorphous materials; the consequences of these plastic deformations are usually referred to as the ``process zone" whose properties one usually does not dare to probe too closely. The aim of this Letter is to do just the opposite \cite{06BPP}. We set up the simplest (from the point of view of symmetries) example of free boundary dynamics, i.e. that of a circular cavity responding to radially symmetric stresses at infinity (or equivalently high pressure inside the cavity). This problem has a rather long history \cite{cavitationRefs, 07BLLP}; the main contribution of our study is the elucidation of the role of amorphous plasticity and the detailed interaction between elasticity and plasticity, throughout the system and in particular near the free boundary where plastic dynamics is explicitly described in terms of the athermal Shear Transformation Zone theory (STZ) \cite{98FL, 07BLP}. This theory automatically includes hardening and
rate-dependent effects in addition to a proper Eulerian description of the equations of motion which
allows the discussion of inertial effects and of large deformations including accelerating and catastrophic cavitation instability.

Consider then an infinite 2D medium
with a circular hole around the origin of radius $R(t)$, loaded by a radially
symmetric stress $\sigma^{\infty}$ at infinity. The symmetry dictates
a radial velocity field $\B v$ that is
independent of the azimuthal angle $\theta$, i.e.
$v_r(\B r,t)\!=\!v(r,t)$ and $v_\theta(\B r,t)\! =\! 0.$
This velocity field implies the components of the total rate of deformation tensor
\begin{equation}
D^{tot}_{rr} \equiv \partial_r v \ , \quad D^{tot}_{\theta\theta}\equiv v/r \ . \label{Dtot}
\end{equation}
In this Letter we restrict the elastic part of the deformation to be small, allowing us to decompose the total rate of deformation tensor into its elastic and plastic parts.
Denoting the material time derivative as ${\cal D}_t \!=\!
{\partial_t}\! + \!v {\partial_r}$,
\begin{equation}
D^{\rm tot}_{ij}  = {\cal D}_t \epsilon^{el}_{ij}+D^{pl}_{ij} \label{elpl} \ ,
\end{equation}
where the plastic part $D^{pl}_{ij} \label{el_pl}$ will be discussed below. The (linear) elastic part
$\epsilon^{el}_{ij}$ in the present symmetry is determined by  the deviatoric stress $ s\!\!\equiv\!\! s_{\theta \theta}\!\!=\!\!-s_{rr}\!\!=\!\! (\sigma_{\theta
\theta}\!-\!\sigma_{rr})/2$ and the pressure $p \!\equiv \!-\!(\sigma_{\theta \theta}\!+\!\sigma_{rr})/2$ according to
 \begin{eqnarray}
\epsilon^{el}_{rr}&=&-p/2K-s/2\mu\ , \nonumber\\
\epsilon^{el}_{\theta
\theta}&=&-p/2K+s/2\mu\label{Hooke} \ ,
\end{eqnarray}
where $\sigma_{ij}$ is the stress tensor and $K$ and $\mu$ are the 2D bulk and shear
moduli respectively.
The plastic rate of deformation is assumed traceless, corresponding to incompressible plasticity.
In the present symmetry the plastic rate of deformation tensor has only one independent component,
$D^{pl}_{rr}\!=\!-D^{pl}_{\theta\theta}\!\equiv\! -D^{pl}$. Substituting now Eqs. (\ref{Dtot}) and (\ref{Hooke}) in
Eq. (\ref{elpl}) one ends up with
\begin{eqnarray}
{\partial_r v} &=&
-{\cal D}_t p/2K-{\cal D}_t s/2\mu-{D^{pl}} \ , \label{kinematic1} \\
v/r&=& -{\cal D}_t p/2K+{\cal D}_t s/2\mu+{D^{pl}} \ . \label{kinematic2} \end{eqnarray}

At this point we introduce the equations of motion for the material density $\rho$ and the velocity $v$ in the form
\begin{eqnarray}
{\cal D}_t \rho &=& -\rho \B\nabla \cdot \B v \label{continuity}\ , \\
\rho{\cal D}_t v&=&\B\nabla \cdot \B \sigma =-
\partial_r \left (r^2 s \right)/r^2 -
\partial_r p \ , \label{EOM}
\end{eqnarray}
where the boundary itself evolves according to $\dot R(t)\! =\! v(R,t)$.
The boundary conditions are given by
\begin{eqnarray}
\sigma_{rr}(R,t)&=&-p(R,t)-s(R,t)=0 \ ,\nonumber\\
\sigma_{rr}(\infty,t)&=&-p(\infty,t)-s(\infty,t)=\sigma^{\infty} \label{BC0} \ .
\end{eqnarray}
As initial conditions for the dynamics we take the static linear elastic solution for a medium
with the given stress at infinity and a hole of radius $R(0)$, assuming that the initial elastic response is much faster than the plastic deformation time scale,
\begin{equation}
p(r,0)=-\sigma^{\infty}\ , \quad
s(r,0)= \sigma^{\infty}R^2(0)/r^2\ ,  \quad v(r,0)=0 \label{initial0} \ .
\end{equation}

Finally, we supply the constitutive relations determining $D^{pl}$. The STZ theory determines this object
in terms of a low density $\Lambda$ of shear transformation zones; these are localized regions embedded in the elastic matrix causing an irreversible deformation in response to shear stress. Under such a stress they are assumed to transform between two internal states, resulting in a plastic strain increment along one of the two principal axes in our symmetry. Macroscopic plastic flow, denoted by $D^{pl}$, results from the average flippings of STZ's. This flow, by itself, can also create and annihilate other STZ's.
The bias of the populations of STZ between these two states, caused by the application of a deviatoric stress, is denoted by a tensor $\B m$ which in
our symmetry has only one independent component denoted by $m$. This field is a local order parameter acting as a ``back stress''. The total density of STZ's is affected by a normalized effective temperature $\chi$, characterizing the state of configurational disorder of the material, via a Boltzmann-like factor. Details of the derivation of the equations can be found in \cite{07BLP}. Here we just state the final equations where $s$ had been normalized with the dynamic yield stress $s_y$, which is a material parameter,
\begin{eqnarray}
\label{eq:Dpl0}
\tau_0 {D^{pl}} &=& \epsilon_0 \Lambda C(s)\left({\rm sgn}(s)-m\right) \ , \\
\label{eq:m0}
{\cal D}_t m &=&
\frac {2D^{pl}}{\epsilon_0 \Lambda}\left(1-\frac{s~m ~e^{-1/\chi}}{\Lambda}\right) \ ,\\
\label{eq:Lambda0}
{\cal D}_t \Lambda&=&
\frac {2sD^{pl}}{\epsilon_0 \Lambda}\left(e^{-1/\chi}- \Lambda\right) \ ,\\
\label{eq:chi0}
 c_0{\cal D}_t \chi &=& 2sD^{pl} (\chi_\infty - \chi) \ ,
\end{eqnarray}
with
\begin{equation}
{\cal C}(
s)=\,\frac{\zeta^{\zeta+1}}{\zeta!}\int_0^{|
s|}(| s|- s_{\alpha})\,
s_{\alpha}^{\zeta}\,\exp (-\zeta \, s_{\alpha})\,d
s_{\alpha}\label{integral0}\ .
\end{equation}
The parameter $\tau_0$ sets the basic time scale of plasticity. $c_0$ and $\epsilon_0$ are parameters of the
order of unity that in this work are taken as unity. The function ${\cal C}(s)$ quantifies the stress-dependence of the transition rate of individual STZ's between their internal states, parameterized here by $\zeta$ that describes the width of STZ's transition thresholds distribution. $\chi_\infty$ is the long time limiting value of the effective temperature \cite{07BLP}.
Note that the assumption that $\Lambda$ is small results in a separation of time scales between
the fast variables $\Lambda$ and $m$ and the slow variable $\chi$. An important feature of the constitutive equations is that the onset of {\em homogenous} unbounded plastic flow results from an exchange of dynamic stability of the bistable field $m$, cf. Eq. (\ref{eq:m0}). For $s\!\le\!1$, i.e. a deviatoric stress below $s_y$, the stable fixed-point is $m\!=\!\pm {\rm sgn}(s)$, corresponding to jamming, $D^{pl}\!=\!0$. On the other hand, for $s\!>\!1$, the stable fixed-point is $m\!=\!1/s$, corresponding to flow $D^{pl}\!\ne\!0$ (with $\Lambda\!=\!e^{-1/\chi}$). This observation justifies our previous choice of normalizing stresses by the dynamic yield stress $s_y$. The consequences of the exchange of stability in the homogeneous equations is further discussed below in relation to the cavitation instability in our highly inhomogeneous configuration.

The full set of equations was solved numerically. To avoid dealing with an infinite time-dependent domain we applied the following time-dependent coordinate transformation
\begin{equation}
\xi=R(t)/r \ . \label{trans}
\end{equation}
This transformation allows us to integrate the equations in the time-independent finite domain $\xi\! \in\! [0,1]$. The price is that new terms are generated in the equations. This domain was discretized uniformly with 301 nodes.
Controlling the equations at small distance required the introduction of an artificial viscosity on the RHS of Eq. (\ref{EOM}). The term introduced is $\rho\eta\! \nabla^2\! v$, with $\eta$ chosen of the order of the square of
space discretization over the time discretization. Time and length are measured in units of $\tau_0$ and $R(0)$ respectively. $\Lambda$ and $m$ are set initially $(t\!=\!0)$ to their respective fixed-points.

\begin{figure}
\centering
\epsfig{width=.45\textwidth,file=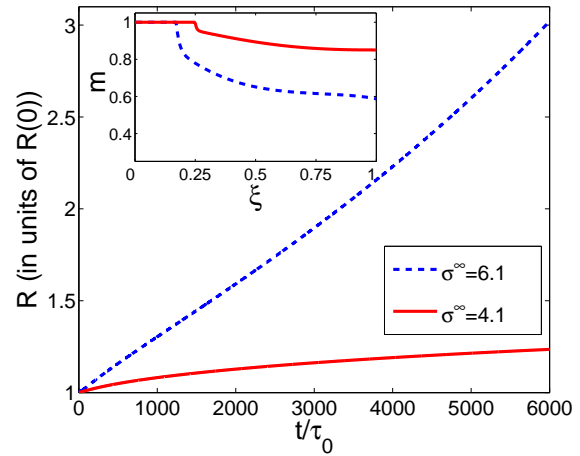}
\caption{(Color online) The time-dependent radius of the cavity for asymptotically stable (solid line)
and asymptotically unstable growth (dashed line). Here and below the parameters of the model are
$\chi_{\infty}\!=\!0.13$, $ \chi(0)\!=\!0.1$,
$\mu\!=\!50$, $\rho=1$, $\zeta\!=\!15$. In these runs $K=100$. (Inset) $m$ as a function of $\xi$ at $t=10^4\tau_0$ for the two cases. For the bounded growth (solid line), $m$ near the free boundary $\xi=1$ increases towards the jamming fixed-point $m\!=\!1$, while in the unstable growth (dashed line) $m$ is at the flowing fixed-point $m\!=\!1/s$ with $s\!>\!1$. Wherever $m\!\ne1$ there exists a plastic flow, while where $m=1$ the material is jammed.}
\label{cavinst}
\end{figure}

The first result to be discussed is the existence of a cavitation instability, such that for
$\sigma^\infty$ below a threshold value $\sigma^{th}$ the circular hole attains an asymptotic stable radius,
whereas for $\sigma^\infty\!> \!\sigma^{th}$ the growth of the cavity is unbounded, signaling a catastrophic
failure. The respective dynamics are demonstrated in Fig. \ref{cavinst}.

The value of $\sigma^{th}$ can be estimated as shown in the appendix of Ref. \cite{07BLLP}, by considering
the quasi-static and incompressible analog of the present problem. For the parameters at hand (except for $K\!=\!\infty$ and $\zeta\!=\!10$) this procedure yielded the estimate $\sigma^{th}\!\approx\! 5$ in agreement
with the present results. For ductile materials such stresses are hard to achieve and indeed cavitation instabilities were observed in the past in ductile materials reinforced by or bonded to brittle materials \cite{experiments}.
To understand the existence of instability, note that in typical situations the stress at infinity
results first in a rapid elastic relaxation to the equilibrium solution of Eq. (\ref{initial0}); this is how the level of applied stress is transmitted to the free boundary. When the deviatoric stress exceeds the material yield stress $s_y$, plastic flow initiates, expanding the cavity beyond the radius of elastic equilibrium. During this process energy is dissipated and the stress tends to relax towards $s_y$. When the expansion is slow, the stress indeed relaxes to $s_y$, the material becomes jammed and the growth is bounded. However, when $\sigma^\infty$ is sufficiently large the rate of growth is such that plastic relaxation fails to reduce the level of stress below $s_y$, resulting in a cavitation instability. This discussion is highlighted by the inset of Fig. \ref{cavinst} where the profiles of $m$ are presented.

\begin{figure}
\centering
\epsfig{width=.45\textwidth,file=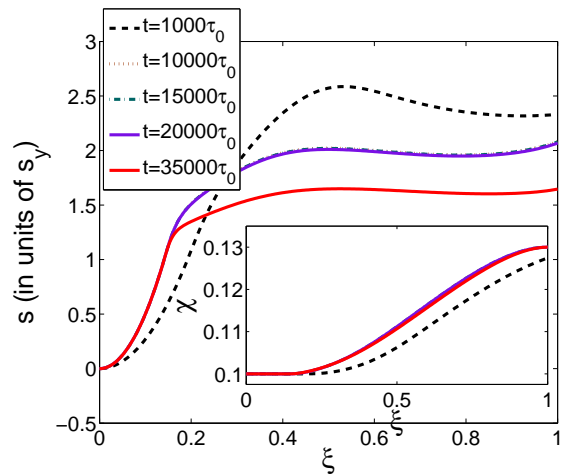}
\caption{(Color online) The deviatoric stress as a function of the variable $\xi$ at different times.
Note the transient short time, the regime of self similarity and the break down of self-similarity at long times. (Inset) The effective temperature $\chi(\xi)$. Note the indistinguishable curves in the self-similar regime. In fact, even after the breakdown of self-similarity $\chi$ does not change significantly. In the plastically deforming region $s\!>\!1$ and $\chi\!>\chi(0)$ (plastic deformation produces disorder). Here the plastic region extends over $R(t)\!<\!r\!<\!6.25R(t)$ ($0.16\!<\xi\!<1$).}
\label{s}
\end{figure}

One learns considerable amount of the physics of the problem by analyzing the velocity of the
free boundary in the unstable phase. This velocity has three typical regimes; the first is a transient
in which $s$ relaxes from its initial value $\sigma^\infty$ to a typical value somewhat larger than $s_y$.
The second regime is very interesting, characterized by the fact that the radius $R(t)$ is the only
typical scale in the problem. We thus expect the velocity of the free boundary to satisfy an
equation of the form
\begin{equation}
\dot R/R = \omega(\sigma^\infty)>0 \ , \label{dotR}
\end{equation}
where $\omega$ is positive and time-independent in this regime. This exponential growth is
accompanied with a self-similar solution in all the fields of the problem, which can depend
only on $\xi$, cf. Eq. (\ref{trans}). As an example we present in Fig. \ref{s} the deviatoric stress $s(\xi)$
and the effective temperature $\chi(\xi)$ for different times, including the onset of the third regime in which self-similar solutions break down
due to inertial effects.
This last regime manifests itself when the velocity of the free boundary
becomes a finite fraction of the typical elastic wave speed; when this happens we have
a new length scale that can be made from the typical wave speed times time, breaking the
self similarity. In order to estimate the velocity in which inertia becomes important we note
that at low velocities almost all the elastic energy release goes to plastic dissipation (below we show
that the external work at infinity is negligible). At high velocities we expect that a finite fraction
of the elastic energy release translates into kinetic energy. We thus want to
compare the kinetic energy density of the medium with the plastic dissipation density. The typical kinetic energy density near the free boundary is given by
$\rho \dot R^2\! /2 $. The plastic dissipation density is given by
$sD^{pl} \tau_0 e^{1/\chi}$. Consulting Eq. (\ref{eq:Dpl0}), one observes that near the free boundary $D^{pl}\tau_0 e^{1/\chi}$ is of the order of unity and $s\! \gtrsim\! s_y$.
Therefore the ratio of the two energy densities is of the order of unity when
\begin{equation}
\label{inertia_import}
\dot R \sim \sqrt{s_y/\rho} \sim c_s \sqrt{s_y/\mu} \ .
\end{equation}
Thus in this problem inertia becomes important at smaller velocities than in the context of brittle fracture where the estimate is $v\! \sim\! c_s$ \cite{98Fre} (recall that $\sqrt{s_y/\mu}$ is significantly smaller than unity for most materials).
\begin{figure}
\centering
\epsfig{width=.45\textwidth,file=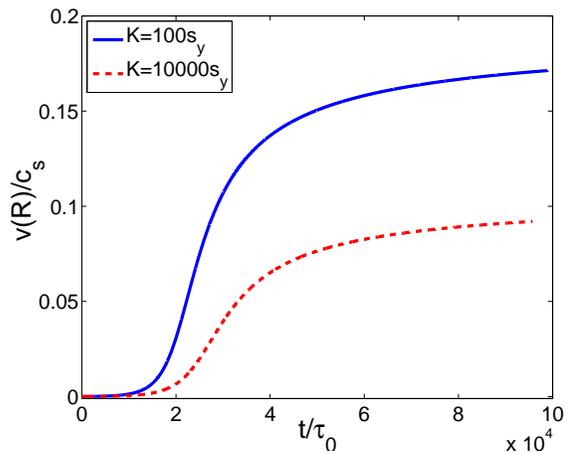}
\caption{(Color online) The velocity of the cavity as function of time, for two different values
of the bulk modulus $K$.}
\label{velocity}
\end{figure}
\begin{figure}
\centering
\epsfig{width=.45\textwidth,file=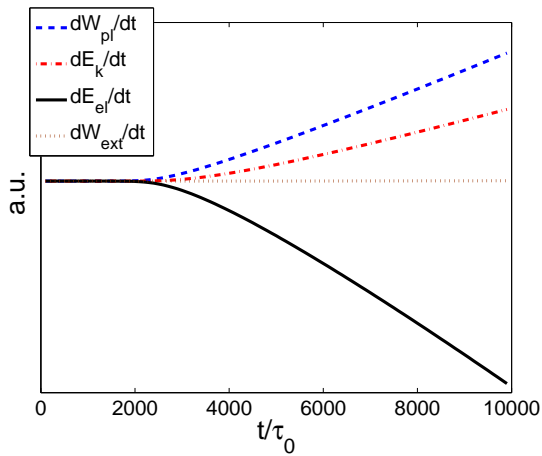}
\caption{(Color online) The rate of change of the various energies in the problem as a function of time.}\label{energ}
\end{figure}
In Fig. \ref{velocity} we present the velocity of the free boundary as a function of time for two values of the bulk modulus $K$. Observe the transient regime, the exponential growth regime, where the problem
exhibits self-similar solutions, and finally the attainment of an asymptotic velocity when inertial effects
become important. In our simulations $\sqrt{s_y/\mu}\!\approx\! 0.14$, and indeed the cross-over for $K\!=\!100$ from the self-similar to the inertial-dominated regime occurs where Eq. (\ref{inertia_import}) predicts.
Increasing $K$ by two order of magnitude (i.e. approaching the incompressible limit) results
in a further decrease in the cross-over point and in the asymptotic velocity. Qualitatively one understands this as the effect of faster longitudinal waves whose velocity $\sim\!\sqrt{K/\rho}$ that are able to ``steal'' energy from the process near the free boundary.

Finally, it is worthwhile to consider the energy exchange mechanisms in this problem. Note that in the standard approach of Linear Elasticity Fracture Mechanics energy balance is used to set up the ``equation of motion" for a crack \cite{98Fre}. On the contrary here we solved the boundary value problem, including the explicit evolution of the free boundary, and it is interesting to go back with this solution at hand to examine the energy exchange. Energy comes in four types: the elastic strain energy $E_{el}$, the kinetic energy $E_K$, the plastic dissipation $W_p$ and the external work $W_{ext}$. We are mainly interested in the time rate of change of these quantities; these are given as the following spatial integrals over the domain $[R(t),\! \bar{R}(t)]$  (the limit $\bar{R} \to \infty$ is briefly discussed below)
\begin{eqnarray}
\label{energies}
\dot{E}_{el}&=&\pi\frac{d}{dt} \int_{R(t)}^{\bar{R}(t)}\epsilon_{ij}^{el}\sigma_{ij}rdr \ ,\quad
\dot{E}_K=\pi\frac{d}{dt} \int_{R(t)}^{\bar{R}(t)}\rho v^2rdr, \nonumber\\
\dot{W}_{ext}&=&2\pi\sigma^{\infty}\dot{\bar{R}}(t)\bar{R}(t) \  , \quad
\dot{W}_{pl}= 4\pi\int_{R(t)}^{\bar{R}(t)} s D^{pl} r dr \ .
\end{eqnarray}
Conservation of energy implies that
\begin{equation}
\dot{W}_{ext}=\dot{W}_{pl}+\dot{E}_{el}+\dot{E}_K \ .
\label{conservation}
\end{equation}
The limit $\bar{R}\!\to\!\infty$ is not well posed as it results in various divergent behaviors. We expect the leading term of $v$ and $\dot{v}$ at $\infty$ to be $\!\sim\! r^{-1}$ (even for a finite $K$). In that case both $E_K$ and $\dot{E}_K$ diverge. Moreover, $E_{el}$ diverges as well for a finite $K$ since $p$ approaches a constant at $\infty$. These are all peculiarities of 2D that should not affect the basic physics we describe.

In Fig. \ref{energ} we plot the different rates of change of energy appearing in Eqs. (\ref{energies}), integrated
up to $300\! R(t)$. $\dot{W}_{ext}$ equals the sum of the three other time derivatives such that Eq. (\ref{conservation}) is satisfied. Note also that $\dot{W}_{ext}$ is significantly smaller than the other three time derivatives. Under this condition, the Fig. \ref{energ} demonstrates explicitly how the release of elastic energy drives the ``crack'', converting this energy into plastic dissipation and kinetic energy.

One question beyond the scope of this Letter is the shape stability of the perfectly symmetric cavity. This issue will be picked-up in a forthcoming publication. In addition, the STZ approach should be applied to cracks in less symmetric situations. This requires significant investment in tensorial generalizations
and in boundary tracking algorithms, which are yet other issues under study.

This work is supported in part by the Minerva Foundation, Munich, Germany, by the Israel Science Foundation and the German-Israeli Foundation. E.B is supported in part by the Center for Complexity Science.

\end{document}